\documentclass[aps,prl,twocolumn,showpacs,superscriptaddress]{revtex4}
\usepackage{graphicx}

\begin{document}

\title{Elementary Excitations in One-Dimensional Electromechanical Systems; \\
Transport with Back-Reaction}

\author{Kang-Hun Ahn}
\affiliation{ Department of Physics, Chungnam National University, Daejon 305-764, Korea}
\author{Hangmo Yi}
\affiliation{ School of Physics, Korea Institute for Advanced Study, Seoul 130-722, Korea }
\date{ \today}

\begin{abstract}
Using an exactly solvable model, we study low-energy properties of a one-dimensional
 spinless electron fluid contained in a quantum-mechanically moving wire located in
a static magnetic field.
The phonon and electric current are coupled via Lorentz force and 
the eigenmodes are described by two independent boson fluids. 
At low energies, the two boson modes are charged while 
one of them has excitation gap due to back-reaction of the Lorentz force.
The theory is illustrated by evaluating optical absorption spectra.
Our results are exact and show a non-perturbative regime of electron transport.
\end{abstract}
\maketitle

Electronic and mechanical properties of nano-electromechanical
systems (NEMS) are recently of fundamental and technical interest\cite{bishop}.  
Perpetual trend of miniaturization of electronic devices 
has already led NEMS smaller down to molecular scale, e.g., nanotube electromechanical systems\cite{kim}
which are
likely the bases of new novel devices in NEMS application\cite{rueckes}.
These systems are close to the regime of quantum-mechanical operation
which is the ultimate limit of nanoelectromechanical devices.
For the design of quantum mechanically operating NEMS devices, it is crucial 
to understand first the quantum effects of mechanical motion coupled to electric current.
Here we ask ourselves;
Are the electric currents influenced by the quantum nature of mechanical motion?
If so, what is the signature of the quantum fluctuation?

To answer these, we study one-dimensional spinless electron fluids coupled to their quantum-mechanical motion
in a static magnetic field. 
This model is exactly solvable as will be shown here.
In this way, we could describe both the electrical and mechanical degrees of freedom 
in  a purely quantum mechanical manner on an equal footing.
The harmonic-fluid approach used for electron fluids in a static wire, which 
are referred to as (Tomonaga-)Luttinger liquid theory\cite{tomonaga,luttinger,haldane}, 
could be successfully applied to
 describe the electric currents coupled to mechanical motions through Lorentz force. 
This magneto-electro-mechanical coupling is essential in so-called electromotive-force technique which 
is popularly used in NEMS-based force sensors\cite{cleland,mohanty}. 

In this paper, we show that electric currents in 
nano-electromechanical wires are strongly 
influenced by the quantum fluctuations of mechanical motion through the magneto-electro-mechanical coupling.
This coupling causes interesting and peculiar phenomena  because of
the feedback reaction between electric and mechanical movement.
As we shall see, 
an important frequency scale related to a gap structure arises due to this feedback reaction. 
The feature we found could be observed
generally in other vibrating one-dimensional electronic systems 
such as atomic wires\cite{chain} and quasi-one-dimensional metals\cite{density}.

We model the electrons with an interacting spinless Fermion fluid with a periodic boundary condition $\Psi(x+L)=\Psi(x)$.
These Fermions are contained in a wire of mass density $\rho_{m}$ and
tension ${\mathcal T}$, and
a magnetic field ${\bf B}$ is applied perpendicular to x-axis. 
The displacement $u(x)$ is in the direction of  $\hat{\bf x} \times {\bf B}$, and
assumed small ($\partial_{x}u \ll 1 $) and periodic ($u(x+L)=u(x)$).
The electrons and the wire are assumed strongly tightened so any force exerted on the electrons in u-direction is 
delivered to the wire immediately. 
The Hamiltonian  reads ($\hbar=1$, $e < 0$)
\begin{eqnarray}
\nonumber
H&=&\frac{1}{2m}\int dx \Psi^{\dagger}(x)\big( \partial_{x} -i e A_{x} \big)^{2}\Psi(x)
\\ 
&+&\frac{1}{2} \int dx dx^{\prime}V(x-x^{\prime}) \rho (x) \rho (x^{\prime})+H_{u}
\\ \nonumber
H_{u}&=&\frac{1}{2}\int dx \big[ \frac{1}{\rho_{m}}\pi_{u}(x)^{2}+{\mathcal T}(\partial_{x} u(x))^{2}  \big].
\label{hamiltonian}
\end{eqnarray}
Here $\pi_{u}$ is the conjugate momentum of $u(x)$; $[u(x),\pi_{u}(x^{\prime})]=i\delta(x-x^{\prime})$.
$A_{x}=Bu(x)+\Phi_{B}/L$ is the gauge potential for the 
applied magnetic field where $\Phi_{B}$ is a constant magnetic flux 
enclosed by the circle of radius $L/2\pi$.
For the sake of convenience, let us redefine $u(x)+\Phi_{B}/BL\rightarrow u(x)$ so that $A_{x}=Bu(x)$.

Following the approach of Haldane\cite{haldane},
we use  a Boson field $\theta$ 
defined from the Fermion density
 $\rho(x)=  (k_{F}+\partial_{x} \theta)/\pi$  
and its conjugate field $\phi$ ( $[\phi(x) , \theta(x^{\prime})]=i\frac{1}{2}\pi {\rm sgn}(x-x^{\prime}) )$.
The boson fields describe the long wavelength fluctuations of the Fermion field $\Psi^{\dagger}(x)$.
It proves convenient to define new Fermion field
 $\psi^{\dagger}=\exp(i\Lambda)\Psi^{\dagger}$, where
$\Lambda(x)=eB\int^{x}u(x^{\prime})dx^{\prime}$.
In terms of $\theta$,$\phi$ and $\Lambda$, the new Fermion field $\psi^{\dagger}$ is written as
\begin{eqnarray}
\psi^{\dagger}(x)\sim \sum_{\rm n~ odd}\exp[ i n [\theta(x)+k_{F}x ]  ] \exp[ i(\phi(x)+\Lambda(x))]
\label{newfield}
\end{eqnarray}
The boundary conditions for the Boson fields are $\theta(x+L)=\theta(x)+\pi {\rm N}$, 
$\phi(x+L)=\phi(x)+\pi {\rm J}$, and $\Lambda(x+L)=\Lambda(x)+eBu_{0}L$, where 
$u_{0} = \frac{1}{L}\int_{0}^{L}dx u(x) $, and
the sum of the topological numbers N and J 
is an even integer. 
Note that the new Fermion field now has a twisted boundary condition $\psi(x+L)=\exp(-ieBu_{0}L)\psi(x)$.

The effective Hamiltonian describing long-distance physics is given by
taking terms of $n=\pm 1$ in Eq.(\ref{newfield}) with a proper normalization constant;
\begin{eqnarray}
H=\frac{v}{2\pi}\int dx \big( g \big(\partial_{x}\phi +eBu \big)^{2}
+g^{-1}\big(\partial_{x}\theta\big)^{2}\big)+H_{u}.
\label{hamiltonian-boson}
\end{eqnarray}
Here the dimensionless constant $g$ is governed by 
electron-electron interaction strength and $v=v_{F}/g$ is the electronic wave velocity, where
$v_{F}$ is the Fermi velocity for the non-interacting case.
For short-ranged electron interactions $V(x-x^{\prime})=V\delta(x-x^{\prime})$, 
$g$ is given by $1/\sqrt{1+V/\pi v_{F}}$.
In terms of the boson fields, the electrical current density is given by 
\begin{eqnarray}
J(x)=-ev_{F}\partial_{x}\phi(x)-v_{F}e^{2}Bu(x).
\label{current}
\end{eqnarray}

Note that the term of $u^{2}$ in Eq.(\ref{hamiltonian-boson}) and $H_{u}$ describe phonons
of finite mass at small momenta;
\begin{eqnarray}
\frac{gv}{2\pi} e^{2}B^{2}\int dx u^{2}(x)+H_{u} 
=\sum_{q}\tilde{\omega}_{s}(q)a^{\dagger}_{q}a_{q},
\label{Hph}
\end{eqnarray}
Here 
\begin{eqnarray}
\tilde{\omega}_{s}(q)=\sqrt{v^{2}_{s}q^{2}+\omega^{2}_{B}},
\end{eqnarray}
where
$v_{s}=\sqrt{ {\mathcal T}/\rho_{m}  }$ is the bare sound velocity and 
 $\omega_{B}=\sqrt{ \frac{v_{F}e^{2}} {\pi\rho_{m}}   }B$.
The first term in Eq.(\ref{Hph}) essentially gives mass to phonon and originates from
{\it back-reaction process} in the system. 
Note that when the wire moves, 
an induced electric field ${\mathcal E}_{\rm emf}=-B\partial_{t}u$ changes 
the momentum of electrons by 
$\int dt e{\mathcal E}_{\rm emf}=-eBu$,
which causes the change in electric current
$-e^{2}Bu\rho_{0}/m$ where  $\rho_{0}=k_{F}/\pi$ is the mean Fermion density.
The Lorentz force $ -v_{F}e^{2}B^{2}u/\pi$   
exerted on the induced electric current  again acts on the movement of the
wire, which we call here the back-reaction process.

 Now let us use the following Fourier-transformed field operators;
\begin{eqnarray}
\nonumber
\theta(x)&=&\theta_{0}+\frac{\pi{\rm N}x}{L}-i\sum_{q\neq 0}
\sqrt{\frac{g\pi}{2|q|L} } (b^{\dagger}_{q}+b_{-q}){\rm sgn}(q) e^{iqx} \\
\nonumber
\phi(x)&=&\phi_{0}+\frac{\pi {\rm J}x}{L}-i\sum_{q\neq 0}
\sqrt{\frac{\pi}{2g|q|L} } (b^{\dagger}_{q}-b_{-q}) e^{iqx} \\
\nonumber
u(x)&=&u_{0}+\sum_{q\neq 0}
\sqrt{\frac{1}{2\rho_{m}\tilde{\omega}_{s}(q)L} } (a^{\dagger}_{-q}+a_{q}) e^{iqx} \\
\pi_{u}(x)&=&\frac{\pi_{u 0}}{L}+i\sum_{q\neq 0}
\sqrt{\frac{\rho_{m}\tilde{\omega}_{s}(q)}{2L}  } (a^{\dagger}_{-q}-a_{q}) e^{iqx}. 
\end{eqnarray}
Here ($\theta_{0}$,J) and ($\phi_{0}$,N) are pairs of conjugate action-angle variables\cite{haldane} and
$[u_{0},\pi_{u0}]=i$.

Using these operators, the Hamiltonian in Eq.(\ref{hamiltonian-boson}) is written as
\begin{eqnarray}
\nonumber
H&=&\sum_{q}v|q|b^{\dagger}_{q}b_{q}+\sum_{q}\tilde{\omega}_{s}(q)a^{\dagger}_{q}a_{q}+H_{ZM}\\
&+&\frac{\omega_{B}}{2}\sum_{q}\sqrt{\frac{v|q|}{\tilde{\omega}_{s}(q)}}
(b_{-q}^{\dagger}-b_{q})(a_{q}^{\dagger}+a_{-q}){\rm sgn}(q)
\label{hamilt-op}
\end{eqnarray}
Here $H_{ZM}$ describes the zero-mode;
$H_{ZM}=\frac{\pi v_{F}}{2L}({\rm J}+\frac{eBu_{0}L}{\pi})^{2}+ \frac{\pi_{u 0}^{2}}{2\rho_{m}L} 
+\frac{v\pi}{2gL}{\rm N}^{2}$,
according to which the center-of-mass of the wire executes 
a simple harmonic oscillation with resonance frequency $\omega_{B}$.

\begin{figure}
\includegraphics[width=7cm,height=5.5cm]{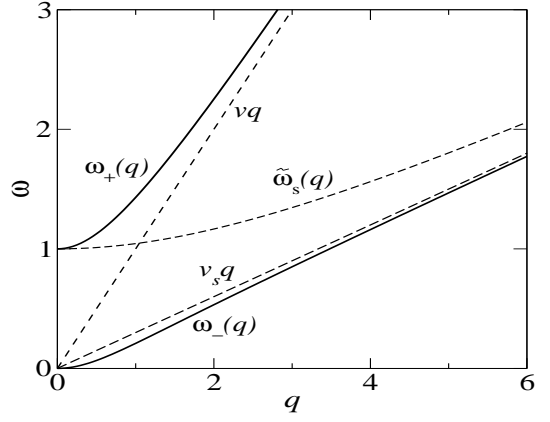}
 \caption{ 
The dispersion relation of the eigenmodes $\omega_{\pm}(q)$ (solid lines).
$\omega$ is in unit of  $\omega_{B}$ and $q$ is in unit of $\omega_{B}/v$.
The dotted lines denote the boson mode for electron $\omega=vq$, phonon $\omega=v_{s}q$,
 and the massive phonon $\tilde{\omega}_{s}(q)$.
We chose $v_{s}=0.3v$ in this figure.  } 
\end{figure}

The Hamiltonian in Eq.(\ref{hamilt-op} )is now diagonalized using new Boson operators
$\gamma_{q, \pm}$, 
which are linear combinations of the four boson operators $b_{q},b_{-q}^{\dagger},a_{q}$, and $a_{-q}^{\dagger}$;
\begin{eqnarray}
H=\sum_{q,\pm}\omega_{\pm}(q)\gamma^{\dagger}_{q, \pm}\gamma_{q, \pm}+H_{ZM}
\end{eqnarray}
where the dispersion relations for the two eigenmodes are 
\begin{eqnarray}
&&\omega_{\pm}(q)= \\
\nonumber
&&\sqrt{ \frac{ (v^{2}+v_{s}^{2})q^{2}+\omega_{B}^{2} \pm
\sqrt{ ((v^{2}-v_{s}^{2})q^{2}-\omega_{B}^{2} )^{2}+4\omega_{B}^{2}v^{2}q^{2} } } {2} }.
\end{eqnarray}
Now the ground state is the vacuum of new bosons, $\gamma_{q \pm}|0>=0$.

For detailed discussion, let us assume $v > v_{s} $ which is usually the case.
In Fig. 1, we plot dispersion relations for the two modes. 
As $q$ increases or $B$ decreases, the system goes to a weak coupling regime where
 the $`+'$ mode  becomes electronic ($\gamma_{q +} \sim b_{q}$,  $\omega_{+}(q) \sim vq$) and 
the $'-'$ mode becomes phononic ($\gamma_{q -} \sim a_{q}$, $\omega_{-}(q) \sim \tilde{\omega}_{s}(q)$).
In the opposite limit where $q$ is small and $B\neq 0$, the system is in a
strong coupling regime where 
 the two uncoupled modes undergo level crossing as shown in Fig. 1 and eventually completely exchange their identity 
in $q\rightarrow 0$ limit,
i.e., $\gamma_{q,+} \rightarrow a_{q}$ and $\gamma_{q,-}\rightarrow b_{q}$.
It is instructive to note that electric current can flow 
through the '+' modes even in $q\rightarrow 0$ limit, where the '+' modes become phononic.
This is because of the 'diamagnetic' contribution of
electric current  $-v_{F}e^{2}Bu(x)$ in Eq.(\ref{current}).

In the strong coupling regime, the dispersion relation of the '+' mode is in the same form with
 plasmon dispersion relation, 
$\omega_{+}(q) \sim \sqrt{v^{2}q^{2}+\omega_{B}^{2}}$ ($q \ll \omega_{B}/v$).
In contrast to the plasmon\cite{mahan}, however, the frequency gap $\omega_{B}$  originates from
the back-reaction which gives mass to phonon at long-wavelength limit,
$\tilde{\omega}_{s}(q) \rightarrow \omega_{B}$. 
The '$-$' mode is gapless  $\omega_{-}(q)\sim v_{s}vq^{2}/\omega_{B}$ ($q \ll \omega_{B}/v_{s}$). 
This mode is also charged at small $q$, 
where electronic and phononic excitation are entangled with each other. 
These excitations at long wavelengths are non-perturbative and can not be captured by perturbative
expansions in $B$.

Now let us consider experiments where the mechanical quantum fluctuations can be observed. 
In Fig. 2, we depict a system we consider.
An external electrical field $V_{e}(t)/d$ is applied along the wire in $0<x<d$ by nearby external metallic
gates with oscillating voltage difference $V_{e}(t)=V_{e}(\omega)e^{i\omega t}$.
If the current $I(\omega)$ is measured far away from the regime $0<x<d$, 
the current will not be sensitive to the position of the contacts so can be given by its averaged value 
 $I(\omega) \approx \frac{1}{d}\int_{0}^{d}dx J(x,\omega)$. 
From the linear response theory, one gets  
\begin{eqnarray}
\alpha(\omega)&\equiv&\frac{I(\omega)}{V_{e}(\omega)}=\frac{1}{i\omega d^{2}}\int_{0}^{d}dx\int_{0}^{d}dx^{\prime} \Pi(x,x^{\prime},\omega),
\label{gw}
\end{eqnarray}
where
  $\Pi(x,x^{\prime}\omega)=i\int_{0}^{\infty}ds < [ J(x,0),J(x^{\prime},-s)]>e^{i\omega s}$ is the current-current
correlation function.

This form is equivalent to the conductance formula $G(\omega)$ used by Kane and Fisher\cite{kane} 
but caution is needed here.
In our case,  additional electromotive force $V_{\rm emf}(\omega)$ is induced along the wire when $B \neq 0$.
In usual two-terminal conductance measurements, since the measured voltage $V(\omega)$ is given by
$V_{\rm emf}(\omega)+V_{e}(\omega)$, the two-terminal conductance is related to our $\alpha(\omega)$
as $I(\omega)=G(\omega)(V_{e}(\omega)+V_{\rm emf}(\omega))=\alpha(\omega)V_{e}(\omega)$.

The real part of $\alpha(\omega)$ should be understood as an optical absorption spectrum.
After averaging over a time period $2\pi/\omega$,
the power $P(\omega)$ 
absorbed from the external electric field to the wire
 is given by $P(\omega)=\frac{1}{2}{\rm Re}\int_{0}^{d}dx (V^{*}_{e}(\omega)/d)J(x,\omega) 
=\frac{1}{2} {\rm Re}\alpha(\omega)|V_{e}(\omega)|^{2}$.

\begin{figure}
\includegraphics[width=8.0cm,height=5cm]{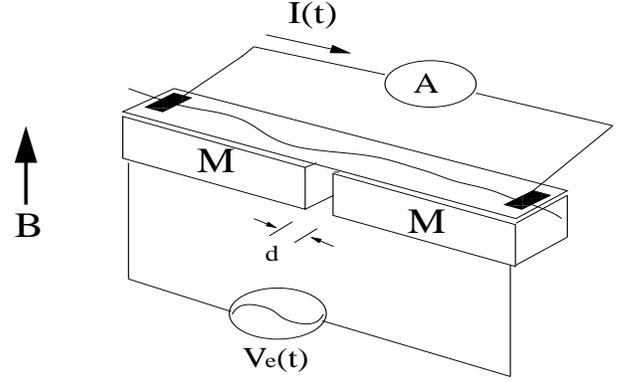}
 \caption{ 
Schematic figure for the experimental set-up in consideration. 
The wire is on an insulating layer above two metallic gates denoted by M.  } 
\end{figure}

To evaluate conveniently the current correlation function, we switch to a Lagrangian formalism
following Kane and Fisher\cite{kane}.
The Hamiltonian in Eq.(\ref{hamiltonian-boson}) is equivalent to the following Euclidean action,
\begin{eqnarray}
\nonumber
S&=& 
\frac{v}{2g} \int d\tau dx 
\left( (\partial_{x} \theta)^{2} 
+\frac{1}{v^{2}}(\partial_{\tau} \theta )^{2} \right)
\\
\label{action}
&+& \frac{\mathcal T}{2} \int d\tau dx
\left( (\partial_{x} u)^{2} 
+\frac{1}{v_{s}^{2}}(\partial_{\tau} u )^{2} \right)
\\ \nonumber
&-& B
\int d\tau dx J(x,\tau) u(x,\tau),
\end{eqnarray}
where $J(x,\tau)=ie\partial_{\tau}\theta(x,\tau)/\pi$.
Evaluation of the current correlation function 
$\Pi(x,x^{\prime},i\omega_{n})$
at Matsubara frequencies $\omega_{n}=2\pi n/\beta$
can be easily done using the above Euclidean action;
\begin{eqnarray} 
\nonumber
&&\Pi(x,x^{\prime},i\omega_{n})=\int_{0}^{\beta}d\tau e^{-i\omega_{n}\tau}<{\rm T}_{\tau}J(x,\tau)J(x^{\prime},0)> \\
            &=&
-\frac{e^{2}v_{F}}{2\pi^{2}}\int_{-\infty}^{\infty}dq e^{iq(x-x^{\prime})}
\frac{\omega_{n}^{2}} 
{ q^{2}v^{2} +\omega_{n}^{2} 
+\omega_{B}^{2} \frac{\omega_{n}^{2}}{q^{2}v_{s}^{2} +\omega_{n}^{2}} }.
\end{eqnarray}
Now, by performing analytic continuation
  $\Pi(x,x^{\prime},\omega)=\Pi(x,x^{\prime},i\omega_{n} \rightarrow \omega +0^{+})$,
we get ${\rm Re} \alpha(\omega)$ from Eq.(\ref{gw});
\begin{eqnarray}
\nonumber
&&{\rm Re}\alpha(\omega)
\\ \nonumber
&=&
\frac{e^{2}}{4\pi} v_{F}  |\omega|
\int_{-\infty}^{\infty}dq \frac{\sin^{2}(qd/2)}{(qd/2)^{2}}
\frac{ \omega^{2}-(qv_{s})^{2}} {\omega^{2}-(\omega_{+}^{2}(q)+\omega_{-}^{2}(q))/2}
\\ &\times&
\left\{ 
\delta(\omega^{2}-\omega_{+}^{2}(q))
+\delta(\omega^{2}-\omega_{-}^{2}(q))
\right\}.
\end{eqnarray}
The result of the above integration is written
in terms of dimensionless parameters  $\tilde{\omega}=|\omega|d/v_{s}$,
$\tilde{B}=L \omega_{B}/v_{s}$, and $\eta=v_{s}/v$; 
${\rm Re} \alpha(\omega)=g\frac{e^{2}}{h}\sum_{\pm}A_{\pm}(\tilde{\omega},\tilde{B},\eta)$.
 (Here we revived the Planck constant.)
The dimensionless function $A_{\pm}$  are 
\begin{eqnarray}
A_{\pm}=\pm \frac{\sin^{2}(\kappa_{\pm}/2)}{(\kappa_{\pm}/2)^{2}}
\frac{\eta(\tilde{\omega}^{2}-\kappa_{\pm}^{2})/\kappa_{\pm}}{\sqrt{(1-\eta^{2})^{2}\tilde{\omega}^{2}+4\tilde{B}^{2}} },
\end{eqnarray}
where $
\kappa_{\pm}=\sqrt{
\left(
(1+\eta^{2})\tilde{\omega}^{2}
\mp\tilde{\omega}\sqrt{(1-\eta^{2})^{2}\tilde{\omega}^{2}+4\tilde{B}^{2}}
\right)/2 }.$

\begin{figure}
\includegraphics[width=7.5cm,height=6cm]{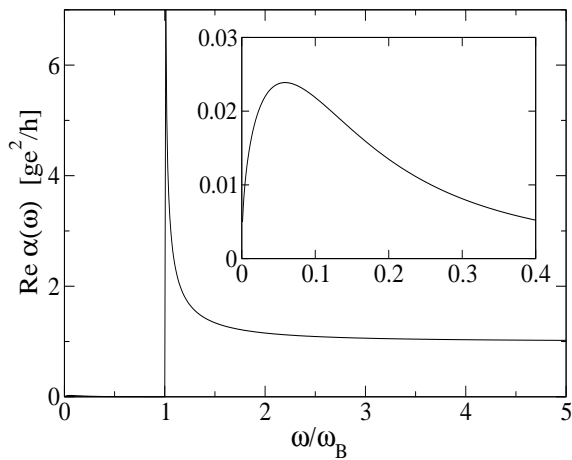}
 \caption{ 
The optical absorption spectrum  for $v_{s}/v=0.1$
 and $\omega_{B}d/v=10^{-4}$.
The inset shows the details at low frequencies.
In the low frequency regime (inset), the  maximum arises near 
$\omega \approx (v_{s}/v)\omega_{B}$. } 
\end{figure}

In Fig. 3, we plot the field absorption power ${\rm Re} \alpha (\omega)$ using the exact formula we obtained.
Since usually $v_{s} \ll v $ 
(for metallic carbon nanotubes, 
 $v_{s}\sim 15$ km/s $\ll v$ and $v \sim 10^{5}$ km/s ), let us assume $\eta \ll 1$.
In this case, the induced electric current
is strongly suppressed when $|\omega| < \omega_{B} $. 
There are three interesting frequency ranges.
(i)~ $\omega \ll  (v_{s}/v)\omega_{B}$. 
Here only '-' modes are excited which are mainly electronic density
fluctuations.
In this regime, using
$A_{-} \approx \frac{\eta}{2}\sqrt{\frac{\tilde{\omega}}{\tilde{B}}} \ll 1$ and $A_{+}=0$,
we get
\begin{eqnarray}
{\rm Re}\alpha(\omega) \approx g\frac{e^{2}}{h}\frac{\sqrt{v_{s}/v}}{2}
\sqrt{\frac{\omega}{\omega_{B}}}
\end{eqnarray}
It is interesting to note the anomalous frequency and magnetic-field dependence 
${\rm Re}\alpha(\omega)\propto \sqrt{\omega/B}$
which can not be obtained within a perturbation analysis.
(ii)~  $(v_{s}/v) \omega_{B} \ll \omega < \omega_{B}$. 
Here also only the '$-$' modes are excited while their nature becomes partially phonon-like
as the frequency increases. 
 In this regime, $A_{-} \approx 4\eta \tilde{B}^{2} \sin^{2}(\tilde{\omega}/2)$
and $A_{+}=0$.
In particular for $\tilde{\omega} < 1 $ (i.e., $\omega< v_{s}/L$), we get
\begin{eqnarray}
{\rm Re}\alpha(\omega) \approx g\frac{e^{2}}{h}(\frac{v_{s}}{v})^{3}(\frac{\omega_{B}}{\omega})^{2}
\end{eqnarray}
(iii)~ Finally when $\omega_{B} < \omega $,
the '+' modes play a significant role and $A_{+}$ has singularity near $\omega=\omega_{B}$.
Near this singular point, in spite of the phononic nature of the '+' modes, 
significant electric current may flow through its 'diamagnetic' contribution;
\begin{eqnarray}
{\rm Re}\alpha(\omega) \approx g\frac{e^{2}}{h}\frac{|\omega|}
{\sqrt{\omega^{2}-\omega_{B}^{2}}} 
\end{eqnarray}
The singularity of $\alpha(\omega)$ originates from the density of states of the 
massive boson mode $\omega_{+}(q)$ near $\omega=\omega_{B}$.
Similar to Landauer conductance quantization,
a quantization rule ${\rm Re}\alpha(\omega) \approx ge^{2}/h$ appears in weak coupling regime 
$\omega_{B} \ll \omega < v/d $. 

Important to all the physical consequences discussed in this work is
a characteristic frequency $\omega_{B}$ associated with the back-reaction 
\begin{eqnarray}
\omega_{B}=\sqrt{\frac{v_{F} e^{2} } {\rho_{m} \hbar \pi}} B
=0.88\times 10^{-2}B[{\rm T}]\sqrt{\frac{v_{F}[m/s]}{\rho_{m}[kg/m]}}
~{\rm Hz},
\label{AYfrequency}
\end{eqnarray}
where we recovered $\hbar$.
As clear in its form, electron-electron interactions do not affect this characteristic frequency.
For single-walled carbon nanotubes, 
the linear mass density is estimated to be $\rho_{m}\sim 4 \times 10^{-13}$ kg/m when its radius is $\sim$ 4 nm .
Using a typical Fermi velocity $v_{F}=10^{5} m/s$, 
we estimate $ \omega_{B} \approx  4.4 ~B[{\rm T}] ~{\rm MHz} $, which is
in the experimentally accessible range. 

While we consider electro-mechanical coupling via Lorentz force,  
capacitive coupling to external gates\cite{kim, sapmaz} might be of interest in other contexts. 
In this case, the coupled charge density and phonon displacement is 
described by $\sim \int dx u(x)\partial_{x}\theta$. 
We assumed no external source of mechanical energy 
dissipation of the wire,
which is another subtle issue of electromechanical system\cite{ahn} 
and important to clarify 
the detailed operating conditions of quantum NEMS.

In summary, we outlined a theory for electrons in one-dimensional wires in
quantum mechanical motion.
Quantum mechanical displacement coupled to electric current
and the back reaction from electrons 
give rise to a current-carrying boson mode with an energy gap.
The induced  electric current 
by oscillating  electric field might reveal the elementary excitations in the system
through its frequency dependence. 
We presented here a non-perturbative regime where the induced electric current
has non-trivial frequency dependence.

\begin{acknowledgements}
This work was supported by Korea Reseacrh Foundation (KRF-2003-070-C00020).
We thank the partial financial support of BK21 physics research division 
in Seoul National University where a part of this work was done (K.H.A.) and the
Swiss-Korean Outstanding Research Efforts Award Program (H.Y.). 
\end{acknowledgements}

\end{document}